\documentclass[useAMS,usenatbib,referee]{biom}
\usepackage{url}
\usepackage{arydshln}
\usepackage{xcolor}
\usepackage{pifont}
\usepackage{amsmath}
\usepackage{amsfonts}
\usepackage{amssymb}
\usepackage{mathtools}
\usepackage{multirow}
\usepackage{pdfpages}

\def\Htransformation{{g}}
\def\Hparam{{\alpha}}

\def\Stransformation{{f_s}}
\def\Sparam{{\beta_s}}

\def\Rtransformation{{f_r}}
\def\Rparam{{\beta_r}}

\newcommand{\indep}{\perp \!\!\! \perp}
\newcommand{\given}{\,|\,}
\newcommand{\E}{\mathbb{E}}
\newcommand{\Pn}{\mathbb{P}_n}

\newtheorem{proposition}{Proposition}

\title[Data integration methods for micro-randomized trials]{Data integration methods for micro-randomized trials}

\author{E. Huch$^{1,*}$\email{ekhuch@umich.edu}, I. Nahum--Shani$^{1,**}$\email{inbal@umich.edu}, L. Potter$^{2}$, C. Lam$^{2}$, D.W. Wetter$^{2}$, and W. Dempsey$^{1,***}$\email{wdem@umich.edu}\\
$^{1}$d3 Center, University of Michigan, Ann Arbor, Michigan, U.S.A. \\
$^{2}$Department of Population Health Sciences and Huntsman Cancer Institute,\\University of Utah, Salt Lake City, Utah, U.S.A.}

\begin{document}


\date{{\it Received March} 2024.}



\pagerange{Pages} 
\volume{Volume}
\pubyear{2024}
\artmonth{March}


\doi{fake/doi.numbers}


\label{firstpage}


\begin{abstract}
Existing statistical methods for the analysis of micro-randomized trials (MRTs) are designed to estimate causal excursion effects using data from a single MRT.
In practice, however, researchers can often find previous MRTs that employ similar interventions. 
In this paper, we develop data integration methods that capitalize on this additional information, leading to statistical efficiency gains.
To further increase efficiency, we demonstrate how to combine these approaches according to a generalization of multivariate precision weighting that allows for correlation between estimates, and we show that the resulting meta-estimator possesses an asymptotic optimality property.
We illustrate our methods in simulation and in a case study involving two MRTs in the area of smoking cessation.
\end{abstract}

%



\maketitle


%

\section{Introduction}
\label{integration:sec:intro}

Micro-randomized trials \citep[MRTs;][]{liao2016sample, qian2022microrandomized} are becoming an increasingly popular tool for developing just-in-time adaptive interventions (JITAIs) in mobile health (mHealth).
JITAIs leverage digital technologies to adapt the delivery of interventions to the rapidly changing needs of individuals \citep{nahum2015building, nahum2018just}.
To empirically inform these interventions, MRTs involve rapid sequential randomizations, meaning that each participant is randomized many times during the trial with short time intervals (e.g., a few hours or minutes) between subsequent randomizations.
The goal of an MRT is to collect data that will enable investigators to estimate the causal effect of delivering just-in-time interventions on a proximal health-related outcome; e.g., the effect of delivering (vs. not delivering) a mobile-based supportive message on tobacco use in the next hour.

MRT data is typically analyzed with the weighted and centered least squares method \citep[WCLS;][]{boruvka2018assessing}.
WCLS is a semi-parametric method that enables unbiased estimation of time-varying treatment effects even when the model includes time-varying endogenous variables. WCLS has been extended in various directions, including for use with different estimands \citep{dempsey2020stratified}, binary outcomes  \citep{qian2021estimating}, clustered data structure \citep{shi2023assessing}, and greater statistical efficiency via auxiliary variables and supervised machine learning methods \citep{shi2023incorporating, shi2023meta}.
However, WCLS has not yet been extended to allow for data integration across multiple studies.

The benefit of data integration methods is that they can result in substantial increases in statistical efficiency relative to a single-study analysis.
They are particularly useful in causal inference because experimental data sets are often small, which makes statistical efficiency a key consideration \citep{shi2023data}.
The potential for efficiency gains depends largely on the size of the additional data set and its relevance to the target estimand.
Investigators seeking to develop a JITAI often use previous MRTs and/or observational intervention studies as a starting point for new studies.
Consequently, previous studies often contain data that are relevant to estimating the target estimand in the study under consideration, indicating the possibility of efficiency gains via data integration.

The key challenge in realizing these efficiency gains is that the causal effects may differ between studies, so naively pooling the data sets together will, in general, result in biased estimates.
Here, we propose an approach that achieves asymptotic unbiasedness under a weaker assumption---that the causal effects must be equal only in a conditional sense.
We show that this assumption leads to several estimation strategies.
Because the resulting estimators are correlated with one another, the classical method of precision weighting \citep[][Chapter 6]{konstantopoulos2019,hedges2014statistical} is not optimal for combining estimates.
Instead, we introduce a meta-estimator that generalizes precision weighting to allow for correlation both \emph{within} and \emph{between} vector-valued estimates, and we show that it possesses an asymptotic optimality property.

The paper proceeds as follows.
Section \ref{integration:sec:preliminaries} formally introduces the problem setup and WCLS.
Sections \ref{integration:sec:conditional_mean_methods}, \ref{integration:sec:density_ratio_methods}, and \ref{integration:sec:combination_methods} develop our proposed data integration methods.
Sections \ref{integration:sec:simulation} and \ref{integration:sec:case_study} present results from applying our methods in a simulation and case study, respectively.
Section \ref{integration:sec:discussion} closes with a discussion.

\section{Preliminaries}
\label{integration:sec:preliminaries}

This section introduces the problem setup, including the target estimand and basic causal assumptions.
We then provide the main technical details for WCLS and discuss the extension of our problem setup to multiple studies.

\subsection{Single-study problem setup}
\label{integration:sec:setup}

For a given participant, we observe a sequence of covariates ($X_t$), treatment assignments ($A_t$), and outcomes ($Y_t$) for $T$ time points.
To simplify the exposition, we assume that $A_t$ is binary (we discuss the extension to three or more treatment levels in Section \ref{integration:sec:discussion}).
We denote the full treatment history up to time point $t$ as
\[H_t = (X_1, A_1, Y_2, X_2, A_2, Y_3, ..., X_t).\]
We adopt the potential outcomes framework of \citet{rubin1974estimating} and posit the existence of potential values for all variables from $Y_2$ onward in the above sequence.
For $Y_2$, $X_2$, and $A_2$ the potential values are $Y_2(a_1)$, $X_2(a_1)$, and $A_2(a_1)$, respectively, with $a_1 \in \{0, 1\}$. After $A_2$, the potential values depend on multiple treatment assignments.
For instance, the potential value for $Y_3$ is $Y_3(a_1, a_2)$.
To make the notation more compact, we use an overbar to represent the history of a given variable up to a particular time point; e.g., $\bar{a}_3 = (a_1, a_2, a_3)$.
Using this notation, the potential values can then be written as $Y_t(\bar{a}_{t-1})$, $X_t(\bar{a}_{t-1})$, and $A_t(\bar{a}_{t-1})$.
We employ the following standard causal assumptions for dynamic treatment regimes:

\begin{assumption}
\label{integration:assumption:consistency}
 (\emph{Consistency}) The observed variables are equal to the potential variables as follows: $Y_t = Y_t(\bar{A}_{t-1})$, $X_t = X_t(\bar{A}_{t-1})$, and $A_t = A_t(\bar{A}_{t-1})$.
\end{assumption}

\begin{assumption}
\label{integration:assumption:positivity}
(\emph{Strict Positivity}) The treatment assignment probabilities are bounded away from zero and one: $\exists \epsilon \in (0, 0.5]$ such that $\epsilon < \Pr(A_t = 1 \given H_t) < 1 - \epsilon$ for all $t \in [T]$.
\end{assumption}

\begin{assumption}
\label{integration:assumption:sequential_unconfoundedness}
(\emph{Sequential Unconfoundedness}) The treatment assignment at time $t$ is independent of all future potential values; i.e., for all $t \in [T]$ and all $(a_t, a_{t+1}, \dots, a_T) \in \{0, 1\}^{T-t+1}$, we have
\[A_t \indep \big\{Y_{t+1}(\bar{A}_{t-1}, a_{t}), X_{t+1}(\bar{A}_{t-1}, a_{t}), A_{t+1}(\bar{A}_{t-1}, a_{t}), \dots, \phantom{\{}Y_{T+1}(\bar{A}_{t-1}, a_{t}, a_{t+1}, \dots, a_{T})\big\} \big| H_t.\]
\end{assumption}
Because $A_t$ is randomized in an MRT, Assumptions \ref{integration:assumption:positivity} and \ref{integration:assumption:sequential_unconfoundedness} typically hold by design.
The scientific goal is to learn how a set of (potentially) time-varying covariates, $R_t \subset H_t$, moderates the causal effect of $A_t$ on the proximal outcome, $Y_{t+1}$:
\begin{equation}
\label{integration:eq:causal_excursion_effects}
  \E \left\{Y_{t+1}(\bar{A}_{t-1}, 1) - Y_{t+1}(\bar{A}_{t-1}, 0) \given R_t(\bar{A}_{t-1}) = r^* \right\}.
\end{equation}
We denote these causal effects as $\Rparam (t, r^*)$.
\citet{boruvka2018assessing} refer to $\Rparam (t, r^*)$ as `causal excursion effects' because they represent the average effect of following the (stochastic) treatment policy up until $t - 1$ and then taking an `excursion' from the policy.
We use $r^*$ to denote a generic value of $r^*$ from the support of $R_t$.
The definition given in \citet{boruvka2018assessing} allows for lags of more than one time point; e.g., the effect of $A_t$ on $Y_{t+2}$.
The extension to additional lags is straightforward, so we use a single lag to simplify the exposition.
\citet{boruvka2018assessing} show that $\Rparam(t, r^*)$ can be identified from the data as
\[\E \left[ \E\left\{Y_{t+1} \given H_t, A_t=1\right\} - \E\left\{Y_{t+1} \given H_t, A_t=0\right\} \given R_t = r^* \right].\]
Following \citet{boruvka2018assessing}, we assume that $\Rparam (t, r^*)$ is a linear function:

\begin{assumption}
\label{integration:assumption:causal_excursion_linear}
    (\emph{Linear $R_t$-moderated Effects}) $\Rparam (t, r^*) = \Rtransformation(r^*)^{\intercal} \Rparam$ for some known function, $\Rtransformation$, determining feature transformations of $r^*$ and some unknown parameter vector, $\Rparam$.
\end{assumption}

In a slight abuse of notation, we use $\Rparam$ to refer to both (1) the nonparametrically defined estimand and (2) the vector of coefficients that determine it according to Assumption \ref{integration:assumption:causal_excursion_linear}.
In principle, $f_r$ and other feature transformation functions we introduce later could depend on $t$;
however, we suppress the dependence in our notation for brevity.

\subsection{Weighted, centered least squares (WCLS)}
\label{integration:sec:wcls}

This section introduces the WCLS method as a means of faciliating the development of our methods in subsequent sections.
To simplify notation, we use $\Pn$ to denote the empirical expectation operator: $\Pn Z = \frac{1}{n}\sum_{i=1}^n Z_i$.
Using this notation, the WCLS estimator is defined as the solution to the estimating equation $0 = \Pn U(\Hparam, \Rparam)$ with $U(\Hparam, \Rparam)$ equal to
\begin{equation}
\label{integration:eq:wcls_ee}
  \sum_{t=1}^{T} W_t \left[Y_{t+1} - \Htransformation(H_t)^{\intercal} \Hparam - \left\{A_t - p_r(1 \given R_t)\right\} \Rtransformation(R_t)^{\intercal} \Rparam\right] 
  \begin{bmatrix}
    \Htransformation(H_t) \\
    \left\{A_t - p_r(1 \given R_t)\right\} \Rtransformation(R_t)
  \end{bmatrix}.
\end{equation}

In Equation \eqref{integration:eq:wcls_ee}, $W_t = p_r(A_t \given R_t) / p_h(A_t \given H_t)$ is a ratio of conditional treatment assignment probabilities.
$\Htransformation(H_t)$ is a transformation of $H_t$ thought to explain variation in the expected value of the outcome.
$\Hparam$ is a corresponding nuisance parameter that is included as a means of increasing statistical efficiency by `de-noising' the outcome.
Crucially, $\Htransformation(H_t)$ need not correspond with the true conditional expectation of $Y_{t+1}$ given $H_t$.
\citet{boruvka2018assessing} show that the WCLS estimator of $\Rparam$ is consistent and asymptotically Gaussian with a covariance matrix that can be consistently estimated via a sandwich estimator.
These sandwich estimators take the form $\mathbf{B}^{-1} \mathbf{M} \mathbf{B}^{-\intercal}$, where $\mathbf{B} = \Pn\, \partial U(\Hparam, \Rparam)$ and $\mathbf{M} = \Pn\, U(\Hparam, \Rparam) U(\Hparam, \Rparam)^{\intercal}$ \citep{stefanski2002calculus}.

We can still apply WCLS when $p_r$, $p_h$, or both are unknown, provided we know the parametric form (e.g., logistic regression) for the unknown probabilities.
In this case, we estimate the corresponding parameters according to their own estimating equations and plug the solutions into the WCLS estimating equation.
We can propagate the uncertainty due to estimating these parameters by `stacking' all of the estimating equations vertically and forming a joint sandwich estimator \citep[See background material in][]{carroll2006measurement}.
Although the parametric form of $p_h$ must be correctly specified, $p_r$ may be assumed constant without loss of consistency or asymptotic normality.

\subsection{Extension to multiple studies}

We now extend the problem setup above to multiple studies.
We assume that $\Rparam (t, r^*)$ is defined in a single study population, and we refer to this study as the ``internal study.''
We allow access to one or more additional ``external studies.''
Mathematically, we distinguish between these studies using the variable $I$ (for ``internal''), which takes on a value of one if an observation is sampled from the internal study population and zero otherwise.
Similarly, we denote the number of participants as $n_1$ and $n_0$ in the internal and external studies, respectively.
We assume the following relationship between the internal and external studies:

\begin{assumption}
\label{integration:assumption:shared_expectation}
(\emph{Shared $S_t$-moderated Effects}) There exists a set of (potentially) time-varying moderators $S_t \supseteq R_t$ such that the $S_t$-conditional causal excursion effects, $\Sparam (t, s^*)$, are equal between studies for all $s^*$:
\begin{align*}
    &\ 
    \E \left\{Y_{t+1}(\bar{A}_{t-1}, 1) - Y_{t+1}(\bar{A}_{t-1}, 0) \given S_t(\bar{A}_{t-1}) = s^*, I=1\right\}
    \\
    =&\ 
    \E \left\{Y_{t+1}(\bar{A}_{t-1}, 1) - Y_{t+1}(\bar{A}_{t-1}, 0) \given S_t(\bar{A}_{t-1}) = s^*, I=0\right\}.
\end{align*}
\end{assumption}

Assumption \ref{integration:assumption:shared_expectation} effectively assumes that the average causal effects are equal between studies within appropriately defined strata.
The $R_t$-moderated causal excursion effects may still differ between studies;
however, Assumption \ref{integration:assumption:shared_expectation} implies that these differences must be due to differences in the conditional distribution of $S_t$ given $R_t$.

\section{Methods based on conditional mean models}
\label{integration:sec:conditional_mean_methods}

Beginning in this section, we develop five data integration methods that extend WCLS to the multi-study setting.
Table \ref{integration:tab:methods} compares the methods and offers recommendations on when each method is most applicable.
This section introduces the first two methods, A-WCLS and P-WCLS, which rely on models for certain conditional means.

\begin{table}
\centering
\small
\begin{tabular}{rccccl}
\hline
        & \multirow{2}{*}{\parbox{47pt}{$\E(\Stransformation(S_t) | R_t)$ model}}
        & \multirow{2}{*}{\parbox{29pt}{$\Sparam(t, s^*)$ model}}
        & \multirow{2}{*}{\parbox{23pt}{$\omega(S_t)$ model}}
        & \multirow{2}{*}{\parbox{49pt}{Combination method}}
        & \\
        & & & & & Recommended use case \\
\hline
A-WCLS  & \ding{51} &           &           &           & Known structure/constraints for $\E(\Stransformation(S_t) | R_t)$\\
P-WCLS  &           & \ding{51} &           &           & High confidence in $\Sparam(t, s^*)$; low confidence in $\omega(S_t)$\\
ET-WCLS &           &           & \ding{51} &           & Subcomponent in PET-WCLS \\
DR-WCLS &           & \ding{51} & \ding{51} & \ding{51} & Medium confidence in both $\Sparam(t, s^*)$ and $\omega(S_t)$\\
PET-WCLS&           & \ding{51} & \ding{51} & \ding{51} & High confidence in both $\Sparam(t, s^*)$ and $\omega(S_t)$\\
\hline
\end{tabular}
\caption{Summary of proposed methods. The first three columns indicate which models are required for each method.
``Combination method'' indicates which methods combine several individual estimators.
``Recommended use case'' presents recommendations on when each method is particularly appropriate; see Section \ref{integration:sec:discussion} for further discussion.} 
\label{integration:tab:methods}
\end{table}

\subsection{A-WCLS}
\label{integration:sec:awcls}

Our first method is based on the following smoothing identity:
\begin{equation}
\label{integration:eq:smoothinf_identity}
    \Rparam(t, r^*) = \E \left\{ \Sparam(t, S_t) \given R_t = r^*\right\}.
\end{equation}

This identity implies that one strategy to estimate $\Rparam(t, r^*)$ is to first estimate $\Sparam(t, S_t)$ and then appropriately average its value conditional on $R_t$.
To that end, we introduce an additional linearity assumption:

\begin{assumption}
\label{integration:assumption:s_causal_excursion_linear}
        (\emph{Linear $S_t$-moderated Effects}) $\Sparam (t, s^*) = \Stransformation(s^*)^{\intercal} \Sparam$ for some known function, $\Stransformation$, determining feature transformations of $s^*$ and some unknown parameter vector, $\Sparam$.
\end{assumption}

Assumption \ref{integration:assumption:s_causal_excursion_linear} is identical to Assumption \ref{integration:assumption:causal_excursion_linear}, replacing $R$'s with $S$'s.
Given Assumption \ref{integration:assumption:s_causal_excursion_linear}, Equation \eqref{integration:eq:smoothinf_identity} implies that
\begin{equation}
\label{integration:eq:E_gs_beta_s}
    \Rparam(t, r^*) = \E \left\{ \Stransformation(S_t) \given R_t = r^*\right\}^{\intercal} \Sparam.
\end{equation}

From this identity, we observe that it would be sufficient to estimate the expected value of each component of $\Stransformation(S_t)$ conditional on $R_t=r^*$ and multiply the resulting vector by $\Sparam$.
For concreteness, we now assume a simple parametric form for this expectation:

\begin{assumption}
\label{integration:assumption:E_gs_linear}
    (\emph{Linear $\E \left\{ \Stransformation(S_t) \given R_t\right\}$}) $\E \left\{ \Stransformation(S_t) \given R_t = r^*\right\} = \boldsymbol{\Gamma}^{\intercal} \Rtransformation(r^*)$ for some known function, $\Rtransformation$, determining feature transformations of $r^*$ and some partially known parameter matrix, $\boldsymbol{\Gamma}$.
    We further assume that each column of $\boldsymbol{\Gamma}$ is either fully known or fully unknown.
\end{assumption}

Applying Assumption \ref{integration:assumption:E_gs_linear} to Equation \eqref{integration:eq:E_gs_beta_s}, we see that $\Rparam(t, r^*) = \Rtransformation(r^*)^{\intercal} \boldsymbol{\Gamma} \Sparam$ and, consequently, we have $\Rparam = \boldsymbol{\Gamma} \Sparam$.
We name this method \textbf{A}pportioned-WCLS (\textbf{A}-WCLS) because it apportions the $S_t$-conditional effects among the elements of $\Rtransformation(R_t)$ according to the corresponding conditional expectations.
We make the following clarifying remarks regarding Assumption \ref{integration:assumption:E_gs_linear}.

\begin{remark}
\label{integration:remark:awcls_gr}
    The expressions above indicate that Assumptions \ref{integration:assumption:s_causal_excursion_linear} and \ref{integration:assumption:E_gs_linear} actually imply Assumption \ref{integration:assumption:causal_excursion_linear}.
    We list Assumption \ref{integration:assumption:causal_excursion_linear} as a separate assumption because it is required for the other methods.
    A key insight from the above derivation is that the analyst must choose feature transformations of $R_t$ (encoded by $\Rtransformation$) that allow the expectation of $\Stransformation(S_t)$ to be expressed as a linear function.
    This assumption can be checked using standard regression diagnostics and plots.
\end{remark}

\begin{remark}
\label{integration:remark:awcls_link_function}
    Assumption \ref{integration:assumption:E_gs_linear} is not required for valid inference of $\Rparam(t, r^*)$; rather, it is a simplifying assumption. We could, for instance, assume a GLM-style link function for the various components of $\E \left\{ \Stransformation(S_t) \given R_t = r^*\right\}$, and the methods described below would generalize naturally (though, Assumption \ref{integration:assumption:causal_excursion_linear} would no longer hold).
\end{remark}

\begin{remark}
\label{integration:remark:awcls_structure}
    Because $R_t \subseteq S_t$, some columns of $\boldsymbol{\Gamma}$ are known.
    In particular, for each common element of $\Rtransformation(R_t)$ and $\Stransformation(S_t)$, there will be a corresponding column in $\boldsymbol{\Gamma}$ with a single nonzero entry equal to one.
    In a similar fashion, we could assume that other elements in $\boldsymbol{\Gamma}$ are zero and modify the method below accordingly.
\end{remark}

With the parametric identification formula $\Rparam = \boldsymbol{\Gamma} \Sparam$, estimation proceeds as follows:

\begin{enumerate}
    \item Estimate the unknown treatment assignment probabilities $p_h$, $p_s$, and $p_r$ by solving a set of estimating equations (e.g., score functions for logistic regression).
    \item Estimate $\Sparam$ by applying WCLS to the combined data set.
    \item Set $\sigma_r^2(R_t) = p_r(1 | R_t) \left\{1 - p_r(1 | R_t) \right\}$ and solve the following estimating equation for all $j$ corresponding to non-common elements of $\Rtransformation(R_t)$ and $\Stransformation(S_t)$:
\[0 = \mathbb P_n \,I \sum_{t=1}^{T} \sigma_r^2(R_t) \Big[ \big\{\Stransformation(S_t)\big\}_j - \Rtransformation(R_t)^{\intercal} \gamma_j \Big]\Rtransformation(R_t),\]
\end{enumerate}

The final step is a weighted least-squares regression within the internal-study population.
The weights are not required for consistency or asymptotic normality, but they can improve the efficiency of the resulting estimator.

To simplify the inference procedures, we order the elements of $\Stransformation(S_t)$ and $\Rtransformation(R_t)$ such that the common elements appear first and in the same positions.
We denote the number of common elements as $c$ and the dimensions of $\Rtransformation(R_t)$ and $\Stransformation(S_t)$ as $d_r$ and $d_s$, respectively. We form
$\boldsymbol{\hat{\Gamma}}$ as
\[\boldsymbol{\hat{\Gamma}} = 
        \left[
        \begin{array}{c:c:c:c:c}
            \begin{array}{c}
                \mathbf{I}_c \\
                \hdashline
                \mathbf{0}_{(d_r-c),c}  
            \end{array}&
            \hat\gamma_{d_s - c + 1} & \hat\gamma_{d_s - c + 2} & ... & \hat\gamma_{d_s}
        \end{array}
        \right].\]

We then calculate our estimate of $\Rparam$ as $\hat{\beta}_r = \boldsymbol{\hat{\Gamma}} \hat{\beta}_s$.
Letting $\phi = (\Sparam\ \gamma_{d_s-c+1}\  ... \gamma_{d_s})^{\intercal}$ with $\hat \phi$ defined similarly, standard results in M-estimation then guarantee (under mild regularity conditions) that $\sqrt{n}(\hat \phi - \phi) \overset{d}{\longrightarrow} N(0, \boldsymbol{\Sigma}_{\phi})$ with $\boldsymbol{\Sigma}_{\phi}$ consistently estimated via a sandwich estimator.
The delta method gives the asymptotic distribution of $\hat \Rparam$: $\sqrt{n}(\hat \Rparam - \Rparam)
    \overset{d}{\longrightarrow}
    N(0, \mathbf{D}_{\phi} \boldsymbol{\Sigma}_{\phi} \mathbf{D}_{\phi}^{\intercal}),$
where $\mathbf{D}_{\phi}$ is the total derivative of $\Rparam = \boldsymbol{\Gamma} \Sparam$ with respect to $\phi$:
\[\mathbf{D}_{\phi} =
    \frac{\partial}{\partial \phi^{\intercal}} \Rparam =
    \big[\begin{array}{c:c}
        \boldsymbol{\Gamma} &
        \beta^{\intercal}_{s,(d_s-c+1):d_s} \otimes \mathbf{I}_{d_r}
    \end{array}\big].\]

In practice, we plug in estimates for $\mathbf{D}_{\phi}$ and $\boldsymbol{\Sigma}_{\phi}$, and inference proceeds as usual.

\subsection{P-WCLS}

This section introduces \textbf{P}rojected-WCLS (\textbf{P}-WCLS).
P-WCLS requires Assumptions \ref{integration:assumption:consistency}--\ref{integration:assumption:s_causal_excursion_linear}, but it does not require Assumption \ref{integration:eq:pwcls_projection};
hence, the required assumptions are somewhat weaker than those of A-WCLS.
P-WCLS estimates $\Rparam$ by projecting $\Stransformation(S_t)^{\intercal} \hat{\beta}_s$ onto $\Rtransformation(R_t)$.
In terms of implementation, we estimate $p_h$, $p_s$, $p_r$, and $\Sparam$ as detailed in Section \ref{integration:sec:awcls},
but we replace the estimation of $\boldsymbol{\Gamma}$ with direct estimation of $\Rparam$ according to the following estimating equation:
\begin{equation}
\label{integration:eq:pwcls_projection}
U_{\Rparam} =
    \mathbb P_n \sum_{t=1}^{T}
        I\, \sigma_r^2(R_t)
        \big\{ \Stransformation(S_t)^{\intercal} \hat{\beta}_s - \Rtransformation(R_t)^{\intercal} \Rparam \big\} \Rtransformation(R_t).
\end{equation}

Solving this equation for $\Rparam$ amounts to performing a regression of the estimated $S_t$-moderated causal excursion effects, $\Stransformation(S_t)^{\intercal} \hat{\beta}_s$, on $\Rtransformation(R_t)$ weighted by $\sigma_r^2(R_t)$.
A natural follow-up question is whether and how this estimator differs from A-WCLS.
Theorem \ref{integration:theorem:awcls_pwcls_equivalence} provides a partial answer.

\begin{theorem}
\label{integration:theorem:awcls_pwcls_equivalence}
    Under Assumption \ref{integration:assumption:E_gs_linear}, A-WCLS and P-WCLS are equivalent.
\end{theorem}

The proof relies on straightforward linear algebra and is provided in Web Appendix A.
As pointed out in \cite{stefanski2002calculus}, the estimating equations defining a given estimator are not unique; however, provided two sets of estimating equations produce the same estimator, they also have the same asymptotic variance.
Consequently, the asymptotic for variances for A-WCLS and P-WCLS are also equivalent under Assumption \ref{integration:assumption:E_gs_linear}---a fact that we confirmed empirically while running our simulation study.
We introduce A-WCLS as a separate method because (a) it provides guidance on how to select $\Rtransformation$ (see Remark \ref{integration:remark:awcls_gr}), (b) it is distinct from P-WCLS under modifications of Assumption \ref{integration:assumption:E_gs_linear} (see Remarks \ref{integration:remark:awcls_link_function} and \ref{integration:remark:awcls_structure}), and (c) the interpretation of $\Rparam$ as a linear combination of $\Sparam$ allows us to connect and interpret multiple levels of moderation analysis.
In cases where A-WCLS and P-WCLS are equivalent, we recommend performing inference via P-WCLS because it allows for a simpler implementation---the delta method is not needed.

\section{Methods based on density ratio models}
\label{integration:sec:density_ratio_methods}

This section introduces \textbf{E}xponentially \textbf{T}ilted WCLS (\textbf{ET}-WCLS).
Rather than assuming a model for shared conditional expectations, ET-WCLS relates the internal and external studies via an exponential density ratio model.
We also show how to combine the resulting estimator with WCLS to produce a meta-estimator that is asymptotically superior to WCLS.

\subsection{ET-WCLS}
\label{integration:sec:etwcls}

Similar to A-WCLS and P-WCLS, ET-WCLS assumes that the $S_t$-conditional causal excursion effects are equal between studies (Assumption \ref{integration:assumption:shared_expectation}).
In contrast to A-WCLS and P-WCLS, however, ET-WCLS does not require them to assume a particular parametric form (i.e., Assumptions \ref{integration:assumption:s_causal_excursion_linear} and \ref{integration:assumption:E_gs_linear} do not apply).
Instead, we assume the studies can be related according to a density ratio as follows:

\begin{assumption}
\label{integration:assumption:et_ratio}
    (\emph{Exponential Tilt Density Ratio}) $p(S_t \,|\, I = 0) > 0$ if $p(S_t \,|\, I = 1) > 0$ except, perhaps, on sets of measure zero.
    Further, there exists a set of feature transformations, $d(S_t)$, and corresponding parameter vector, $\omega$, such that the conditional densities of $S_t$ can be related as follows:
\[\omega(S_t)
        \coloneqq \frac
            {p(S_t \,|\, I = 1)}
            {p(S_t \,|\, I = 0)}
        = \exp\left\{ d(S_t)^{\intercal} \omega\right\}.\]
\end{assumption}

Assumption \ref{integration:assumption:et_ratio} allows us to view the external data as an unevenly sampled representation of the internal study population.
This relationship suggests a strategy in which we estimate $\omega(S_t)$ and use it to counterbalance the under- and over-representation of each data point, similar to the methods provided in \citet{graham2012inverse,graham2016efficient} and \citet{tan2006distributional}.
\citet{qin1998inferences} details how to estimate density ratios of this form via a slight modification of logistic regression (due to case-control parity, the non-intercept coefficients are identical).
The full estimation procedure for ET-WCLS is given below:

\begin{enumerate}
    \item If needed, estimate $p_h$ and $p_r$ as explained in Section \ref{integration:sec:awcls}
    \item Maximize the log-likelihood $\ell(\omega) = \Pn \sum_{t=1}^{T} \left( I d(S_t)^{\intercal} \omega - \log\left[1 + \rho \exp\left\{d(S_t)^{\intercal} \omega\right\} \right] \right)$, where $\rho \coloneqq n_1 / n_0$ to obtain an estimate of $\omega$
    \item Estimate $\Rparam$ as the solution to Equation \eqref{integration:eq:wcls_ee}, replacing $W_t$ with $\left[(1 - I)\, \exp\left\{ d(S_t)^{\intercal} \omega\right\}\, W_t\right]$
\end{enumerate}

In Web Appendix A, we prove the following result about ET-WCLS:

\begin{theorem}
\label{integration:theorem:etwcls_estimand}
    Assuming the existence of certain moments, the ET-WCLS estimator is a consistent, asymptotically normal estimator of $\Rparam$.
\end{theorem}

We prove Theorem \ref{integration:theorem:etwcls_estimand} using classical results in M-estimation.
The regularity conditions require the reweighted WCLS equation to have bounded fourth moments.
From a practical standpoint, this requirement means that there must be sufficient overlap between studies in the distributions of $S_t$;
otherwise, ET-WCLS may suffer from instability in a manner similar to other weighting methods, such as inverse probability weighting \citep{robins2007comment, seaman2013review}.

\subsection{Meta-estimator}
\label{integration:sec:meta_estimator}

A common practical challenge with weighting-based estimators is that they can exhibit high variance under insufficient overlap, even if the theoretical requirements are satisfied.
For ET-WCLS in particular, this challenge results in ET-WCLS being less efficient than WCLS-Internal (the single-study analysis) unless the external study is large relative to the internal study.
This dilemma begs the question: Is it possible to pool the WCLS-Internal and ET-WCLS estimates such that the resulting estimator is more efficient than either estimate individually?
This section introduces an asymptotically optimal meta-estimator that provides an affirmative answer to this question.

The central challenge in developing the meta-estimator is that WCLS-Internal and ET-WCLS are, in general, correlated due to estimation of $\omega$, which involves both the internal and external studies.
Consequently, the classical solution of precision weighting is not guaranteed to be optimal.
We note that closely related problems have been considered in the meta-analysis literature.
\citet{konstantopoulos2019_13} summarizes multivariate methods that are applicable to vector-valued effect-size estimates provided there is no correlation between estimates.
In contrast, \citet[Chapter 10]{hedges2014statistical} provide a meta-analysis method for dependent estimates of a scalar-valued parameter.
However, our setting involves both of these challenges: vector-valued estimates with correlation both within \emph{and} between estimates.
We were unable to locate a method that addresses both of these challenges.

The first step toward a solution is to jointly stack all relevant estimating equations.
For full generality, we consider the case with $J \geq 2$ distinct estimators of $\Rparam$.
We stack these estimators in a single vector as follows:
\[\hat{\theta} = \left(\hat{\beta}_{r1}^{\intercal},\ \hat{\beta}_{r2}^{\intercal}, \dots, \hat{\beta}_{rJ}^{\intercal}\right)^{\intercal}.\]

We then form a sandwich estimator, $\hat{\Sigma}$, of $\Sigma \coloneqq \text{Var}\left(\sqrt{n} \hat{\theta}\right)$ that captures the dependence structure both within \textit{and} between the $\hat{\beta}_{rj}$'s.
Letting $\hat{\Lambda} \coloneqq \hat{\Sigma}^{-1}$, the meta-estimator is then defined as

\begin{equation*}
    \hat{\beta}_r
    =
    \left(\sum_{j=1}^J \sum_{k=1}^J \hat{\Lambda}_{j, k}\right)^{-1} \sum_{j=1}^J \hat{\Lambda}_{j.} \hat{\theta}
    =
    (1_J^{\intercal} \otimes \hat{\Omega})\,\hat{\Lambda}\, \hat{\theta}
\end{equation*}

with $\hat{\Omega} = \left(\sum_{j=1}^J \sum_{k=1}^J \hat{\Lambda}_{j, k}\right)^{-1}$.
We denote the population-level analogs of $\hat{\Lambda}$ and $\hat{\Omega}$ as $\Lambda$ and $\Omega$, respectively.
$\hat{\beta}_r$ reduces to standard precision weighting when the off-diagonal blocks of $\hat{\Lambda}$ (or, equivalently, $\hat{\Sigma}$) are zero.
Similarly, it reduces to the estimator given in \citet[Chapter 10]{hedges2014statistical} when $\Rparam$ is a scalar.
$\hat{\Omega}$ is guaranteed to exist and, consequently, $\Rparam$ is well-defined provided $\hat{\Sigma}$ is positive definite.
This property follows from the following proposition, the proof of which is given in Web Appendix A.

\begin{proposition}
\label{integration:propoition:block_matrix_inversion}
    Let $A$ be a positive definite square matrix.
    Then for any partition of this matrix into equal-sized blocks, the sum of these blocks is also positive definite and, hence, invertible.
\end{proposition}

We now turn our attention to the asymptotic properties of $\hat{\beta}_r$.
In Web Appendix A, we prove the following result:

\begin{theorem}
\label{integration:theorem:meta_estimator}
    Consider the class of unbiased estimators of $\Rparam$ that can be written in the form $B \hat{\theta}$ for some matrix $B$ that may depend on $\hat{\Sigma}$.
    Among this class of estimators, $\hat{\beta}_r$ achieves the minimum asymptotic variance, $\frac{1}{n} (1_J^{\intercal} \otimes \Omega) \Lambda (1_J \otimes \Omega)$, which can be consistently estimated as $\frac{1}{n}(1_J^{\intercal} \otimes \hat{\Omega}) \hat{\Lambda} (1_J \otimes \hat{\Omega})$.
\end{theorem}

The proof relies on showing that $\hat{\beta}_r$ is asymptotically equivalent to an oracle estimator having the same mathematical formulation as $\Rparam$, but employing $\Sigma$ and $\Lambda$ instead of $\hat{\Sigma}$ and $\hat{\Lambda}$.
The optimality of this oracle estimator follows from an application of the Gauss--Markov Theorem for generalized least squares.
Theorem \ref{integration:theorem:meta_estimator} immediately implies that pooling the WCLS-Internal and ET-WCLS estimators is asymptotically superior to simply selecting one of the two estimators.
More generally, the meta-estimator provides an asymptotically optimal formula for combining correlated estimates that is applicable in broader settings than those considered here.

One drawback of the meta-estimator is that it relies on accurate estimation of a large, full-rank covariance matrix.
Consequently, we expect its performance to degrade in small samples when this covariance matrix is poorly estimated.
In these settings, one path forward is to assume that $\Sigma$ takes on a Kronecker structure, a common assumption in high-dimensional covariance estimation \citep{srivastava2008models}.
Web Appendix B provides the methodological details.

\section{Combination methods}
\label{integration:sec:combination_methods}

Having developed two general data integration strategies, we now turn our attention to developing methods that combine them for increased robustness and efficiency.

\subsection{DR-WCLS}
\label{integration:sec:dr_methods}

The first combination method we develop is \textbf{D}oubly \textbf{R}obust WCLS (\textbf{DR}-WCLS): a combination of P-WCLS and ET-WCLS that produces valid inferences provided that the assumptions for at least one method hold.
We develop methods separately for the internal and external study, then discuss how to combine them using the meta-estimator of Section \ref{integration:sec:meta_estimator}.
We discuss the internal study first.
Our proposed method is based on a `pseudo-outcome,' $\Tilde{Y}_{t+1}$, similar to those employed in \citet{bang2005doubly} and \citet{kennedy2023towards}:
\begin{equation*}
\begin{aligned}
    &\,\frac
        {Y_{t+1} - m(H_t, A_t)}
        {A_t - p_h(0|H_t)}
      + m(H_t, 1) - m(H_1, 0)
    \\
    =&\frac
        {W_t \{A_t\hspace{-1pt}- p_r(1|R_t)\} \{Y_{t+1}\hspace{-1pt}- m(H_t, A_t)\}}
        {\sigma_r^2(R_t)}
      \hspace{-1pt}+\hspace{-1pt} m(H_t, 1) \hspace{-1pt}-\hspace{-1pt} m(H_1, 0),
\end{aligned}
\end{equation*}

where $m(H_t, A_t)$ is a model having $\E\left\{m(H_t, 1) - m(H_t, 0) | S_t\right\} = \Stransformation(S_t)^{\intercal} \Sparam$, such as
\[m(H_t, A_t)
    = \Htransformation(H_t)^{\intercal} \Hparam + \{A_t - p_s(1 | S_t)\} \Stransformation(S_t)^{\intercal} \Sparam.\]

We then have $E\left[\Tilde{Y}_{t+1} | S_t, I=1\right] = \Sparam(t, S_t)$ \emph{even if} $\Stransformation(S_t)^{\intercal} \Sparam$ is misspecified.
Thus, based on our smoothing identity (Equation \eqref{integration:eq:smoothinf_identity}), we can estimate $\Rparam$ by regressing $\Tilde{Y}_{t+1}$ on $\Rtransformation(R_t)$ in the internal study according to the following estimating equation:
\begin{equation}
\label{integration:eq:internal_dr_equations}
   0 =
    \mathbb P_n I \sum_{t=1}^{T} \sigma_r^2(R_t) \big[ \Tilde{Y}_{t+1} - \Rtransformation(R_t)^{\intercal} \Rparam \big] \Rtransformation(R_t).
\end{equation}

The resulting estimator is fully robust in the sense that it does not require any of Assumptions \ref{integration:assumption:s_causal_excursion_linear}, \ref{integration:assumption:E_gs_linear}, or \ref{integration:assumption:et_ratio} to hold.
However, the estimator above uses the external study only to produce a shared model, $m(H_t, A_t)$, and as such produces limited efficiency gains relative to standard WCLS.

To attain meaningful efficiency gains, we now turn our attention toward developing a doubly-robust estimator that uses the external study more directly.
This estimator cannot be represented using a single pseudo-outcome as above.
Instead, we work directly from an expanded version of Equation \eqref{integration:eq:internal_dr_equations}.
The estimating equation defining the external-study is $0 = \mathbb P_n \sum_{t=1}^{T} \Tilde{U}(\Rparam)$, where $\Tilde{U}(\Rparam)$ is defined as
\begin{equation}
\label{integration:eq:external_dr_equations}
\begin{aligned}
    &\frac{1 - I}{1 - \pi}\, \omega(S_t) W_t \{A_t - p_r(1 | R_t)\} \{Y_{t+1} - m(H_t, A_t)\} \Rtransformation(R_t)
    \\
    &\quad\quad + \frac{I}{\pi}\, \sigma_r^2(R_t) \{m(H_t, 1) - m(H_1, 0)\} \Rtransformation(R_t)
    \\
    &\quad\quad - \frac{I}{\pi}\, \sigma_r^2(R_t)\, \Rtransformation(R_t)^{\intercal} \Rparam \Rtransformation(R_t),
\end{aligned}
\end{equation}

with $\pi$ representing the population proportion of individuals belonging to the internal study.
Note that this nuisance parameter requires its own estimating equation of the form $0 = \mathbb P_n (I - \pi)$.
On the one hand, if the model-based assumptions (Assumptions \ref{integration:assumption:s_causal_excursion_linear} and \ref{integration:assumption:E_gs_linear}) hold, then the first term has expectation zero and the second term (equal to $\Sparam(t, S_t)$ by Assumption \ref{integration:assumption:s_causal_excursion_linear}) is projected onto $\Rtransformation(R_t)$ in a manner similar to P-WCLS.
On the other hand, if Assumption \ref{integration:assumption:et_ratio} holds, then
\[\E \sum_{t=1}^{T} \frac{1 - I}{1 - \pi}\, \omega(S_t) W_t [A_t - p_r(1 | R_t)]\, m(H_t, A_t)
    = \E \sum_{t=1}^{T} \frac{I}{\pi} \sigma_r^2(R_t) [m(H_t, 1) - m(H_t, 0)],\]

so the corresponding terms in Equation \eqref{integration:eq:external_dr_equations} cancel in expectation, leaving only the term involving $Y_{t+1}$ and the last line.
Under Assumption \ref{integration:assumption:et_ratio}, the expectation of the former is $\sigma_r^2(R_t) \Rparam(t, R_t) = \sigma_r^2(R_t) \Rtransformation(R_t)^{\intercal} \Rparam$.
Consequently, $\Rparam$ solves the estimating equation in expectation, which implies that the solution to Equation \eqref{integration:eq:external_dr_equations} is a consistent estimator of $\Rparam$.

Combining these observations, we see that the external study estimator possesses the desired double-robustness property.
After forming these two estimates and the corresponding joint sandwich estimator, we can then combine them via the meta-estimator of Section \ref{integration:sec:meta_estimator}.
The meta-estimator preserves the doubly-robust nature of the individual estimators and, by Theorem \ref{integration:theorem:meta_estimator}, results in an estimator with lower asymptotic variance than either individual estimator.

\subsection{PET-WCLS}
\label{integration:sec:pet_wcls}

In this section, we briefly discuss our second combination method, \textbf{PET}-WCLS, which combines \textbf{P}-WCLS and \textbf{ET}-WCLS, effectively requiring Assumptions \ref{integration:assumption:s_causal_excursion_linear}, \ref{integration:assumption:E_gs_linear}, and \ref{integration:assumption:et_ratio} to hold.
We form the estimator by combining WCLS-Internal, P-WCLS, and ET-WCLS according to our meta-estimator.
When Assumptions \ref{integration:assumption:s_causal_excursion_linear}, \ref{integration:assumption:E_gs_linear}, and \ref{integration:assumption:et_ratio} all hold, PET-WCLS is guaranteed to be asymptotically superior to any other linear combination of its constituent methods by Theorem \ref{integration:theorem:meta_estimator}.

\section{Simulation}
\label{integration:sec:simulation}

This section details a simulation study we performed to empirically evaluate the performance of our proposed estimators.

\subsection{Simulation setup}
\label{integration:sec:simulation_setup}

Throughout the simulation study, we use $T=20$ time points.
The simulation includes two simulated studies: one internal, one external.
We vary the number of participants in each study, $n_1$ and $n_0$, to investigate the impact of sample size on the efficiency of our estimators.
The simulation employs three simulated covariates: $X_{t1}$, $X_{t2}$, and $X_{t3}$.
We set $R_t = X_{t1}$, $S_t = \{X_{t1}, X_{t2}\}$, $\Sparam(t, S_t) = 1 + 2 X_{t1} - 3 X_{t2}$, and $\Rparam(t, R_t) = -2 + 5 X_{t1}$.
Web Appendix E provides additional details on the simulation setup and a link to the computer code.

In the results below, we label methods that use only the internal data set with the suffix -Internal.
The other methods use both the internal and external data sets.
WCLS-Pooled naively applies WCLS to the pooled data set, which generally results in biased estimates.

\subsection{Simulation results}
\label{integration:sec:sim_results}

\begin{table}
\centering
\begin{tabular}{lllllll}
  \hline
\multirow{2}{*}{\parbox{1pt}{Coefficient name}} & \multirow{2}{*}{\parbox{25pt}{True value}} & \multirow{2}{*}{\parbox{1pt}{Method}} & \multirow{2}{*}{\parbox{42pt}{Avg\\estimate}} & \multirow{2}{*}{\parbox{48pt}{Relative\\efficiency}} & \multirow{2}{*}{\parbox{28pt}{rMSE}} & \multirow{2}{*}{\parbox{40pt}{Coverage}} \\ 
   &  &  &  &  &  &  \\
\hline
  \multirow{10}{*}{Intercept} & -2 & WCLS-Internal & \textbf{-1.94} & 100.0\% & 1.57 & \textbf{94.5\%} \\ 
   & -2 & WCLS-Pooled & -0.48 & N/A & 1.77 & 61.0\% \\ 
   & -2 & P-WCLS-Internal & \textbf{-1.94} & 99.9\% & 1.57 & \textbf{95.3\%} \\ 
   & -2 & P-WCLS-Pooled & \textbf{-1.96} & 119.0\% & 1.31 & \textbf{95.8\%} \\ 
   & -2 & ET-WCLS & -2.32 & 113.1\% & 1.42 & \textbf{92.5\%} \\ 
   & -2 & DR-WCLS & -2.27 & 116.6\% & 1.37 & \textbf{95.3\%} \\ 
   & -2 & PET-WCLS & -2.27 & \textbf{136.3\%} & \textbf{1.18} & \textbf{94.3\%} \\ 
 \hline 
  \multirow{10}{*}{Slope} & 5 & WCLS-Internal & \textbf{4.86} & 100.0\% & 1.77 & \textbf{96.0\%} \\ 
   & 5 & WCLS-Pooled & 3.44 & N/A & 1.86 & 58.8\% \\ 
   & 5 & P-WCLS-Internal & \textbf{4.86} & 99.9\% & 1.77 & \textbf{96.0\%} \\ 
   & 5 & P-WCLS-Pooled & \textbf{4.92} & 130.9\% & 1.35 & \textbf{95.8\%} \\ 
   & 5 & ET-WCLS & \textbf{5.02} & 116.7\% & 1.51 & \textbf{93.8\%} \\ 
   & 5 & DR-WCLS & \textbf{5.10} & 124.9\% & 1.42 & \textbf{96.3\%} \\ 
   & 5 & PET-WCLS & \textbf{5.14} & \textbf{153.9\%} & \textbf{1.16} & \textbf{94.0\%} \\ 
   \hline
\end{tabular}
\caption{Results from the simulation with 400 individuals in both the internal and external studies.
For the ``Avg estimate'' and ``Coverage'' columns, the boldface indicates values within Monte Carlo error ($3\sigma$) of the truth.
For the ``Relative efficiency'' and ``rMSE'' columns, the boldface indicates the best performance for each coefficient (PET-WCLS in both cases).} 
\label{integration:tab:simulation_results}
\end{table}

Table \ref{integration:tab:simulation_results} summarizes the performance of each method with $n_1 = n_0 = 400$.
As expected, every method except WCLS-Pooled exhibited minimal bias and near-nominal confidence interval coverage.
We show in Web Appendix F that the bias---albeit minimal---apparent in the density ratio methods (ET-WCLS, DR-WCLS, and PET-WCLS) diminishes with larger sample sizes.

P-WCLS, ET-WCLS, DR-WCLS, and PET-WCLS all outperform WCLS-Internal in terms of statistical efficiency and rMSE.
PET-WCLS performs best---likely because it is asymptotically superior to WCLS, P-WCLS, and ET-WCLS---achieving efficiency improvements of 36.3\% and 53.9\% relative to WCLS-Internal.
P-WCLS outperforms WCLS-Internal only when it uses the external data; otherwise, its performance is quite similar to WCLS-Internal.
Web Appendix F shows additional comparisons, including different sample sizes and method variations.

\section{Case study}
\label{integration:sec:case_study}

In this section, we apply P-WCLS to the MARS and Affective Science studies.

\subsection{Study design}
\label{integration:sec:study_design}
In both studies, the treatment is to send push notifications that prompt participants to engage in evidence-based self-regulatory strategies.
The stated goal of these notifications is to ``improve smokers' ability to resist craving and build self-regulatory skills'' \citep{nahum2021mobile}.

Six times per day for ten days, the mHealth app prompted participants to answer a two-question (2-Q) survey about cigarette availability and current affect.
Immediately upon answering the survey (or about two minutes later in the case of non-response), the app randomly assigned participants to receive a prompt to engage in a self-regulatory strategy with probability 0.5.
Approximately one hour after randomization, the app prompted participants to answer additional questions in an ecological momentary assessment \citep[EMA;][]{shiffman2008ecological}.

\subsection{Data analysis}
\label{integration:sec:case_study_analysis}

At the time of writing, the Affective Science study is still enrolling participants, but the MARS study has completed enrollment and the results are currently being analyzed.
Consequently, we have access only to a subset of preliminary data ($n_1=68$, $n_0=97$) that will be further refined to assess the studies' primary aims.
Although the MARS study study recruited 114 participants, we excluded 17 due to withdrawn consent (3), lack of microrandomizations (1), insufficient EMA adherence (11), or missing baseline tobacco dependence (2).
The main hypotheses for the MARS study are based on self-reported (via EMA) engagement with self-regulatory strategies.
Because our primary goal is to use these studies to demonstrate our methods, we restrict attention to a separate outcome: The amount of time (in seconds) spent in the app activities in the hour following micro-randomization.

To apply our methods, we must select two sets of moderators, $S_t$ and $R_t$.
The goal is to estimate how $R_t$ moderates treatment efficacy.
In contrast, $S_t \supseteq R_t$ is a set of moderators such that Assumption \ref{integration:assumption:shared_expectation} is plausible; i.e., conditional on $S_t$, the causal excursion effects are equal between studies.
We include two variables in $R_t$: an indicator for whether the participant completed the last 2-Q survey and the smoker's baseline tobacco dependence (measured by the participant's response to the question ``How many cigarettes do you smoke per day?'').
$S_t$ includes two additional variables: an indicator for whether the participant identifies as male and the participant's self-reported race/ethnicity.
The race/ethnicity variable is divided into three categories: Latino, non-Latino white, and other, which includes non-Latino black and other minority races.

Based on these choices, Assumption \ref{integration:assumption:shared_expectation} states that participants with the same 2-Q survey completion status, baseline tobacco dependence, gender, and race/ethnicity have the same causal excursion effects regardless of which study they participated in (MARS or Affective Science).
Although we cannot verify this assumption, we can attempt to falsify it via a $\chi^2$ model comparison test.
The resulting p-value is 0.98, indicating essentially no evidence against Assumption \ref{integration:assumption:shared_expectation}.

We apply P-WCLS because (1) it performed well in our simulation study and (2) it is less sensitive to misspecification of its required sub-model ($\beta_s(t, s^*)$) than the methods relying on density ratios.
Table \ref{integration:tab:case_study_coefs} displays the estimates of $\Rparam$ via WCLS-Internal, P-WCLS-Internal, and P-WCLS-Pooled.
As expected, P-WCLS-Pooled produces the smallest standard errors because it leverages the MARS data to improve statistical efficiency.
For completeness, we also show results for our other methods in Web Appendix C.

\begin{figure}
\centering
\includegraphics[width=\columnwidth]{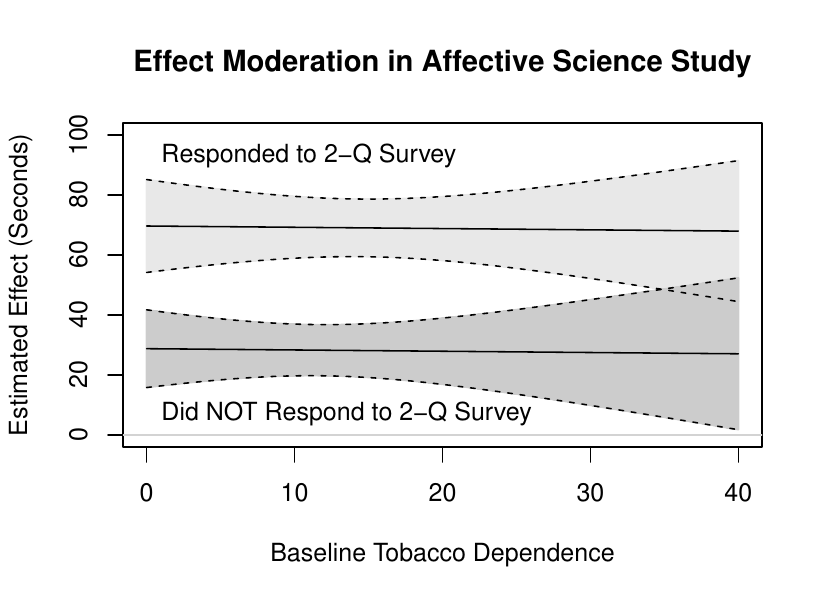}
\caption{Estimated effect of prompts on time spent in the mobile app within the Affective Science study.
Dashed lines denote pointwise 95\% confidence bands.}
\label{integration:fig:case_study_effects}
\end{figure}

Figure \ref{integration:fig:case_study_effects} provides a visual depiction of the estimated effects.
We have strong evidence of a positive effect across most values of $R_t$.
The estimated coefficients suggest that the effects are more than twice as large among those who responded to the previous 2-Q survey compared to those who did not.
In contrast, we see limited evidence that a smoker's baseline tobacco dependence (mean: 13.1, SD: 7.6) moderates these effects.

\begin{table}[ht]
\centering
\begin{tabular}{lrrr}
  \hline
 Coefficient Name & WCLS-Internal & P-WCLS-Internal & P-WCLS-Pooled \\ 
  \hline
Intercept & *25.15 (9.50) & *25.16 (10.04) & *28.77 (\textbf{6.61}) \\ 
  Tobacco Dependence & 0.09 (0.58) & 0.09 (0.61) & -0.04 (\textbf{0.43}) \\ 
  2-Q Survey Completion & *39.47 (7.35) & *39.45 (7.76) & *40.89 (\textbf{6.13}) \\ 
  \hline
  \end{tabular}
\caption{Estimates of $\beta_r$ (standard errors) from applying WCLS-Internal, P-WCLS-Internal, and P-WCLS-Pooled in the case study; asterisks denote statistical significance at the 0.05 level.
P-WCLS-Pooled produces the smallest standard errors.} 
\label{integration:tab:case_study_coefs}
\end{table}

\section{Discussion}
\label{integration:sec:discussion}

This paper presents five methods for pooling data across multiple MRTs for the purpose of estimating causal excursion effects.
The simulations demonstrate that the methods outperform the standard single-study analysis (WCLS-Internal) without losing the ability to perform calibrated inference on $\Rparam$.
The case study provides an illustrative example of (1) the mechanics of applying the method to real MRTs and (2) the potential for efficiency gains.

In practice, researchers must decide which method and corresponding assumptions to employ for a given analysis.
While implementing the case study and simulation, we found that the exponential tilt models were more difficult to specify accurately than the conditional mean models, in part because common regression diagnostics can be used to check the latter.
Moreover, when the exponential tilt models are misspecified, we observed that the resulting estimators---including the robust estimators---can be highly unstable.
Our recommendations in Table \ref{integration:tab:methods} largely reflect this finding.

All five data integration methods rely on Assumption \ref{integration:assumption:shared_expectation}---that the $S_t$-moderated causal excursion effects are equal between studies.
Future work might consider how to relax this assumption, perhaps assuming bounds on discrepancies between the $S_t$-moderated causal excursion effects in each study.
Under this modified assumption, a natural approach would be to penalize the internal-study estimates toward those of the external study.
This approach would require careful consideration of the resulting bias--variance tradeoff.

In addition to the data integration methods, we also developed an asymptotically optimal meta-estimator that generalizes the classical method of precision weighting.
We suspect this meta-estimator could be applied in other settings outside of mHealth, especially in the areas of meta-analysis and data integration.

An interesting direction for future research would be to combine some of the ideas here with other extensions of WCLS;
e.g., the extensions to binary outcomes or clustered MRTs.
This direction would allow the application of these methods in broader mHealth contexts.
\citet{boruvka2018assessing} briefly mention that it is possible to extend WCLS to the setting of three or more treatment levels by introducing multiple centered treatment indicators.
Web Appendix D provides a proof that this strategy does, in fact, result in consistent estimates and discusses how to apply this strategy to our methods.

As these methods are applied to future MRTs, one potential challenge is identifying how to relate measurements and treatments across different study designs.
At a minimum, our proposed methods require (1) treatments with similar effects, (2) a shared set of moderators, and (3) a similar outcome between studies.
Moreover, these variables must be selected so that Assumption \ref{integration:assumption:shared_expectation} is plausible.
Relaxations of Assumption \ref{integration:assumption:shared_expectation} could help address this challenge, but the practical relevance of external studies will remain a central consideration in whether and how to integrate data across MRTs.


\backmatter


\section*{Acknowledgements}

The authors thank Jamie Yap and Qinggang Yu (d3c team members) for their assistance in curating the data for the case study. This work was supported in part by grant P50DA054039 provided through the NIH and NIDDK. The MARS and Affective Science studies were funded by grants U01CA229437 and R01CA224537, respectively, provided through the NIH and NCI.

\section*{Data availability statement}

The data underlying this article cannot be shared publicly due to privacy of the participants involved in the MARS and Affective Science studies.

\vspace*{-8pt}


%

\bibliographystyle{biom} \bibliography{references}






\section*{Supplementary Materials}

Web Appendices referenced in Sections~\ref{integration:sec:conditional_mean_methods}, \ref{integration:sec:density_ratio_methods}, \ref{integration:sec:simulation}, and \ref{integration:sec:case_study} are available at the end of this document.
The code for the simulation study is publicly available at \url{https://github.com/eastonhuch/mrt-data-integration.git}.
\label{lastpage}



\end{document}